\documentclass[aps,pra,twocolumn,superscriptaddress]{revtex4}
\usepackage{bm, amsmath, amsfonts}
\usepackage{graphicx}
\usepackage{color}

\begin{document}

\title{Separable and entangled states in the high-spin XX central spin model}
\author{Ning Wu}
\email{wun1985@gmail.com}
\affiliation{Center for Quantum Technology Research, School of Physics, Beijing Institute of Technology, Beijing 100081, China}
\author{Xi-Wen Guan}
\email{xwe105@wipm.ac.cn}
\affiliation{State Key Laboratory of Magnetic Resonance and Atomic and Molecular Physics, Wuhan Institute of Physics and Mathematics, APM, Chinese Academy of Sciences, Wuhan 430071, China}
\affiliation{Center for Cold Atom Physics, Chinese Academy of Sciences, Wuhan 430071, China}
\affiliation{Department of Theoretical Physics, Research School of Physics and Engineering, Australian National University, Canberra ACT 0200, Australia}
\author{Jon Links}
\email{jrl@maths.uq.edu.au}
\affiliation{School of Mathematics and Physics, The University of Queensland, 4072, Australia}
\begin{abstract}
\par It is shown in a recent preprint [arXiv:2001.10008] that the central spin model with XX-type qubit-bath coupling is integrable for a central spin $s_0=1/2$. Two types of eigenstates, separable states (dark states) and entangled states (bright states) between the central spin and the bath spins, are manifested. In this work, we show by using an operator product state approach that the XX central spin model with central spin $s_0>1/2$ and inhomogeneous coupling is partially solvable. That is, a subset of the eigenstates are obtained by the operator product state ansatz. These are the separable states and those entangled states in the single-spin-excitation subspace with respect to the fully polarized reference state. Due to the high degeneracy of the separable states, the resulting Bethe ansatz equations are found to be non-unique. In the case of $s_0=1/2$ we show that all the separable and entangled states can be written in terms of the operator product states, recovering the results in [arXiv:2001.10008]. Moreover, we also apply our method to the case of homogeneous coupling and derive the corresponding Bethe ansatz equations.
\end{abstract}

\maketitle
\section{Introduction}
\par The central spin model describes a central spin $\vec{S}_0$ interacting with a noninteracting bath composed of $N$ spins $\{\vec{S}_j\}$ via XXZ-type inhomogeneous hyperfine couplings~\cite{Gaudin}. It is described by the Hamiltonian
\begin{eqnarray}\label{Hcsm}
H^{\mathrm{(XXZ)}}_{\mathrm{CSM}}=hS^z_0+\sum^N_{j=1}\left[g_j( S^x_0S^x_j+S^y_0S^y_j)+g'_jS^z_0S^z_j\right],
\end{eqnarray}
where $h$ is an external magnetic field acting on the central spin, and $\{g_j\}$ ($\{g'_j\}$) are the in-plane (Ising) part of the inhomogeneous anisotropic coupling constants. The size of the central spin and the $j$th bath spin are denoted $s_0$ and $s_j$, respectively, each of which can be either an integer or a half odd-integer. With the advent of quantum technologies, the central spin model and related generalizations nowadays play an important role in solid-state based systems, such as electron or hole spins confined in semiconductor quantum dots and Nitrogen Vacancy centers, which are believed to be promising setups to realize quantum computation~\cite{Glazman2002,Glazman2003,Loss2003,RMP2007,Yang-W:2017}. This has stimulated many theoretical studies on both static~\cite{CMP1994,Kiss2001,NPB2005,Faribault2011,Claeys2015,wu2018,Guan2018,XXCSM} and dynamical properties~\cite{Loss2004,Stolze2007,Faribault2013,wu2014,wu2016,Lindoy2018,Guan2019} of central spin systems without/with intrabath coupling. Moreover, the exactly solvable central spin models belong to the class of  Gaudin-type long-range interacting systems \cite{RMP2004}. The existence of exact solutions offers a range of benefits aiding the analysis of Gaudin models including topological properties \cite{Ibanez2009,rombouts10}, efficient numerical procedures \cite{Faribault2011,elaraby12}, benchmarks for undertaking perturbative studies    \cite{claeys17}, and  opportunities for exact calculations in open systems \cite{rowlands18}.
\par For general inhomogeneous and nonzero $\{g_j\}$ and $\{g'_j\}$, it is known that $H^{\mathrm{(XXZ)}}_{\mathrm{CSM}}$ admits exact solutions  through  an \emph{operator product state ansatz}~\cite{wu2018}. These have the structure
\begin{eqnarray}\label{OPSA}
|\psi_M\rangle=\prod^M_{q=1}B^{+/-}_q|\phi\rangle,
\end{eqnarray}
which is generated by acting a set of parameter-dependent collective raising (lowering) operators, $B^\pm_q=\sum^N_{j=0}A_{qj}S^\pm_j$, onto a proper reference state $|\phi\rangle$. Here, $M$ is the number of spin excitations with respect to $|\phi\rangle$ according to the conservation of total magnetization of the system. By employing a similar operator product state ansatz, von Delft and co-workers~\cite{Delft0,Delft2001} studied Richardson's reduced BCS model~\cite{Richardson} and derived the associated Bethe ansatz equations in an elegant and natural way with the help of an \emph{operator approach} based purely on the commutator scheme~\cite{Delft0,Delft2001}. The above method was later successfully applied to the inhomogeneous Dicke model~\cite{Delft2010}, pairing models coupled to a single bosonic mode~\cite{Links2012}, and the anisotropic XXZ central spin model~\cite{wu2018}.
The operator approach has proven to provide a concrete and less abstract tool to treat Gaudin-like models. Throughout this work, a Hamiltonian is said to be solvable (partially solvable) if \emph{all (some) of the eigenstates can be constructed by the operator product state ansatz}. We avoid using the term  \emph{quasi-exactly solvable} here, since this typically applies to an operator defined on an infinite-dimensional vector space which admits an invariant finite-dimensional subspace~\cite{Turb88,Uly92,Ushv93,Dunn08}. The Hamiltonian (\ref{Hcsm}) acts on a finite-dimensional space.
\par It has long been known that $H^{\mathrm{(XXZ)}}_{\mathrm{CSM}}$ is integrable at the isotropic point with $g_j=g'_j,~\forall j$~\cite{Gaudin}. Several types of anisotropic generalizations having specific forms of $\{g_j\}$ and $\{g'_j\}$ are also shown using the Gaudin algebra to be integrable (see \cite{RMP2004} and references therein). Using the aforementioned operator method, it is argued in Ref.~\cite{wu2018} that $H^{\mathrm{(XXZ)}}_{\mathrm{CSM}}$ is solvable for arbitrary spin sizes under the solvability condition
\begin{eqnarray}\label{ggc}
g'^2_j-g^2_j=\mathrm{const},~j=1,2,\cdots,N,
\end{eqnarray}
with general inhomogeneous and nonvanishing couplings. Actually, it is known that for spin-1/2 systems the condition given by Eq.~(\ref{ggc}) can be derived from the Gaudin algebra (see, for example, Refs.~\cite{NPB2005,Claeys2015,Yang:2004,Claeys:2015a}). 
Note that there also exists a class of spin-1/2 XXZ integrable models built from non-skew symmetric $r$-matrices, which do not necessarily obey Eq.~(\ref{ggc}) (see Ref.~\cite{Skrypnyk:2019} for a recent review on the generalized Gaudin models and classification of their corresponding $r$-matrices). In spite of the above-mentioned known results for the XXZ central spin model, less is known about the fully anisotropic limit with $g'_j=0$, where $H^{\mathrm{(XXZ)}}_{\mathrm{CSM}}$ is reduced to the XX central spin model described by $H^{\mathrm{(XX)}}_{\mathrm{CSM}}=hS^z_0+\sum^N_{j=1}g_j( S^x_0S^x_j+S^y_0S^y_j)$. First attempts in this direction were carried out by Jivulescu $et~al$.~\cite{Jivulescu2009}, who employed the procedure proposed in the original paper of Gaudin~\cite{Gaudin} to single out the structure of a subset of exact eigenstates of $H^{\mathrm{(XX)}}_{\mathrm{CSM}}$ for $s_j=1/2,~\forall j$. Most recently, Villazon \emph{et~al}.~\cite{XXCSM} showed by constructing an extensive set of conserved quantities that $H^{\mathrm{(XX)}}_{\mathrm{CSM}}$  is actually integrable for $s_0=1/2$ and real $\{g_j\}$. It is found that the eigenstates of $H^{\mathrm{(XX)}}_{\mathrm{CSM}}$ can be divided into two classes: Dark states having a product state structure between the central spin and bath spins, and bright states for which the central spin is entangled with the spin bath. In this work, the dark states (bright states) will be simply referred to as \emph{separable states} (\emph{entangled states}) due to their different structures.
\par The integrability of $H^{\mathrm{(XX)}}_{\mathrm{CSM}}$ may at first sight seem puzzling, since $g'_j=0$ violates the integrable condition given by Eq.~(\ref{ggc}). However, as shown in Ref.~\cite{wu2018}, condition (\ref{ggc}) is derived under the assumption that the coefficient $A_{q0}$ appearing in $B^\pm_q$ is nonzero for \emph{every} $q$, which is a necessary requirement for $g'_j\neq 0$ in the framework of the operator approach. In this work, we will apply the aforementioned operator approach to the study of the high-spin XX central spin model with $s_0\geq1/2$. Due to the absence of the Ising coupling $g'_j$, the collective raising/lowering operator $B^\pm_q$ in the ansatz (\ref{OPSA}) does not necessarily contain the lowering operator of the central spin, $S^-_0$, giving rise to new solvability conditions other than (\ref{ggc}). It is precisely the number of this kind of new operator (denoted by $Q$ with $0\leq Q\leq M$) in the operator string $\prod_qB^{\pm}_q$ which determines the structure of the eigenstates. Specifically, to guarantee possible (at least partial) solvability of $H^{\mathrm{(XX)}}_{\mathrm{CSM}}$ in the subspace with $M$ spin excitations, it will be shown that $Q$ must be either $M-1$ or $M$.
\par Through a step-by-step construction of the eigenvalue problem based on the operator product approach, we show that for $s_0=1/2$ the case of $Q=M$ ($Q=M-1$) corresponds exactly to the separable (entangled) states revealed in Ref.~\cite{XXCSM}. For $s_0>1/2$, we find that the operator product state ansatz still provides all the separable states with a constant energy, but only a subset of the entangled states lying in the single-spin-excitation subspace. In this sense, the high-spin XX central spin model is only partially solvable. Since the manifold of the separable states is highly degenerate, the Bethe ansatz equations derived do not have a unique form. We finally apply our method to the case of homogeneous coupling. In contrast to the case of inhomogeneous coupling, all values of $Q$ ($0\leq Q\leq M$) are allowed. We also derive the corresponding Bethe ansatz equations.
\par The rest of the paper is organized as follows. In Sec.~\ref{SecII}, we introduce the XX central spin model with complex coupling and the operator approach to be used throughout this work. In Sec.~\ref{SecIII} and Sec.~\ref{SecIV} we study in detail the construction of the separable and entangled states for $Q=M$ and $Q=M-1$, respectively. In Sec.~\ref{SecV} we study the case of the homogeneous coupling and derive the corresponding Bethe ansatz equations. Conclusions are drawn in Sec.~\ref{Secfinal}.
\section{Model and methodology}\label{SecII}
\subsection{The XX central spin model and the operator product state ansatz}
\par We are interested in the XX central spin model described by the Hamiltonian
\begin{eqnarray}\label{Hxx}
H&=&h(S^z_0-s_0)+\frac{1}{2}\sum^N_{j=1}(g_j S^+_0S^-_j+g^*_jS^-_0S^+_j).
\end{eqnarray}
We assume that each $g_j$ is nonvanishing, since otherwise the $j$th bath spin is isolated from the system. The \emph{nonuniform} coupling constants $\{g_j\}$ are allowed to be complex. From the relation $S^x_0S^y_{j}-S^y_0S^x_j=i(S^+_0S^-_j-S^-_0S^+_j)/2$, we can rewrite $H$ as
\begin{eqnarray}\label{HxxDM}
H&=&h(S^z_0-s_0)+\sum^N_{j=1}(\Re g_j)( S^x_0S^x_j+ S^y_0S^y_j)\nonumber\\
&&+ \sum^N_{j=1}(\Im g_j)(S^x_0S^y_j-S^y_0S^x_j),\nonumber
\end{eqnarray}
which shows that the real (imaginary) part of $g_j$ measures the XX-type (Dzyaloshinskii-Moriya-type) interaction between the central spin and the $j$th bath spin. The complex XX-type coupling between the central spin and the bath in Eq.~(\ref{Hxx}) is both interesting and important in various physical scenarios because of its resemblance to the atom-field interaction. For example, it was noted in Refs.~\cite{wu2014,CR2013} that for real and homogeneous coupling the resultant qubit-big spin model can mimic the collapse and revival of Rabi oscillations observed in the Jaynes-Cummings model. Equation (\ref{Hxx}) can also be used to describe the interaction between a bunch of quantum emitters/Frenkel excitons to a single-mode cavity when the cavity mode is approximated as a two-level system in the zero- and single-photon subspaces~\cite{LPP2016}.
\par The $c$-number term $-hs_0$ in Eq.~(\ref{Hxx}) is introduced to make $H$ satisfy $H|F\rangle=0$ and $H|F'\rangle=-2hs_0|F'\rangle$, where
\begin{eqnarray}\label{Fstate}
|F\rangle&=&|s_0\rangle |s_1,\cdots,s_N\rangle,\nonumber\\
|F'\rangle&=&|-s_0\rangle |-s_1,\cdots,-s_N\rangle
\end{eqnarray}
are the highest-weight and lowest-weight state (with the first index denoting the central spin), respectively. Below $|F\rangle$ will mainly be taken as the reference state on which the operator string appearing in the ansatz acts, though nearly equivalent analysis can be performed for $|F'\rangle$. It is easy to see that the total magnetization $\hat{L}_z=\sum^N_{j=0}S^z_j$ (its eigenvalue will be denoted as $L_z$) of the central spin and the spin bath is conserved.
\par To obtain an eigenstate $|\Psi_M\rangle$ of $H$ in the subspace spanned by all spin configurations with magnetization $L_z=\sum^N_{l=0}s_l-M$, we introduce $M$ collective spin lowering operators
\begin{eqnarray}\label{Bq}
B^-_q=\sum^N_{l=0}A_{ql}S^-_l,~q=1,2,\cdots,M
\end{eqnarray}
where $\{A_{ql}\}$ are $M(N+1)$ parameters to be determined by letting the following (unnormalized) operator product state ansatz
\begin{eqnarray}\label{product}
|\Psi_M\rangle=P^M_1|F\rangle,
\end{eqnarray}
satisfy the Schr\"odinger equation
\begin{eqnarray}\label{Sch}
H|\Psi_M\rangle=E_M|\Psi_M\rangle.
\end{eqnarray}
Here,
\begin{equation}
P^n_m\equiv
\begin{cases}
 \prod^n_{q=m}B^-_q&m\leq n,\\
 1&m>n,\\
\end{cases}
	\label{Pmn}
\end{equation}
and $E_M$ is the corresponding eigenenergy. For later use, we also define
\begin{eqnarray}\label{Pmn}
P^{n,(l)}_m&\equiv& P^{l-1}_mS^-_0P^n_{l+1},~(m\leq l\leq n).
\end{eqnarray}
\subsection{The operator approach}
\par Following the operator approach~\cite{wu2018,Delft0,Delft2001}, we start with the identity
\begin{eqnarray}\label{Tr1}
[H,P^M_1]=\sum^M_{q=1}P^{q-1}_1[H,B^-_q]P^M_{q+1},
\end{eqnarray}
which is a direct consequence of the Leibniz rule
\begin{eqnarray}\label{Leibniz}
[x,y_1y_2\cdots y_n]&=&[x,y_1]y_2\cdots y_n+y_1[x,y_2]y_3\cdots y_n\nonumber\\
&&+\cdots+y_1\cdots y_{n-1}[x,y_n]\nonumber
\end{eqnarray}
for arbitrary operators $x,y_1,\cdots$, and $y_n$. The commutator $[H,B^-_q]$ in Eq.~(\ref{Tr1}) can be calculated as
\begin{eqnarray}\label{Cm11}
[H,B^-_q]&=& S^z_0\sum^N_{j=1}A_{q0} g_jS^-_j + S^-_0 \sum^N_{j=1} A_{qj} g^*_jS^z_j\nonumber\\
&&-hA_{q0}S^-_0.
\end{eqnarray}
The above commutator can be simplified by imposing certain constraints on the parameters $\{A_{qj}\}$. The main idea is to gather terms on the right-hand side of Eq.~(\ref{Cm11}) that contain spin lowering operators and demand that their linear combinations take the form of the collective lowering operator $B^-_q$. The usual way is to require $A_{q0} g_j$ to be proportional to $A_{qj}$ with a $j$-independent \emph{nonvanishing} coefficient $-\omega_q$~\cite{wu2018}, i.e.
\begin{eqnarray}\label{Ctr1b}
A_{q0}g_j=-\omega_q A_{qj},
\end{eqnarray}
so that the first term on the right-hand side of Eq.~(\ref{Cm11}) becomes $-\omega_q S^z_0\sum^N_{j=1} A_{qj}S^-_j=-\omega_q S^z_0(B^-_q-A_{q0}S^-_0)$. Note that $A_{q0}$ must be \emph{nonzero} for otherwise we have $A_{qj}=0,~\forall j$.
\par However, there exists an alternative, perhaps more obvious, choice,
\begin{eqnarray}\label{Ctr1a}
A_{q0}=0,
\end{eqnarray}
for which the terms involving $S^-_j$'s in Eq.~(\ref{Cm11}) all vanish, while $A_{qj}$ might be left \emph{arbitrary} at the moment. We emphasize that the choice given by Eq.~(\ref{Ctr1a}) cannot be incorporated into Eq.~(\ref{Ctr1b}) where $A_{q0}$ must be finite.
\par Noting that $[B^-_q,B^-_{q'}]=0$, we can assume without loss of generality that the first $Q$ $(B^-_q)$'s in the ansatz (\ref{product}) are associated with condition (\ref{Ctr1a}), and the remaining $M-Q$
ones are associated with condition (\ref{Ctr1b}):
\begin{eqnarray}
A_{q0}&=&0,~q=1,2,\cdots, Q,\nonumber\\
A_{q0}&\neq& 0,~q=Q+1,Q+2,\cdots, M.\nonumber
\end{eqnarray}
We write
\begin{eqnarray}
\tilde{A}_{qj}=A_{qj},~q=1,\cdots,Q
\end{eqnarray}
to distinguish the $(A_{qj})$'s for $q\leq Q$ from those for $q>Q$, so that
\begin{eqnarray}
\label{Bq1}
B^-_q&=&\sum^N_{j=1}\tilde{A}_{qj}S^-_j,~q=1,\cdots,Q,\nonumber\\
\label{Bq2}
B^-_q&=&\sum^N_{j=0}A_{qj}S^-_j,~q=Q+1,\cdots,M.
\end{eqnarray}
The commutators $[H,B^-_q]$ are accordingly divided into two categories
\begin{eqnarray}
\label{HBq1}
[H,B^-_q]&=&  S^-_0\tilde{X}_q,~q=1,\cdots,Q\\
\label{Xq1}
\tilde{X}_q&\equiv&\sum^N_{j=1} \tilde{A}_{qj} g^*_jS^z_j,
\end{eqnarray}
and
\begin{eqnarray}
\label{HBq2}
[H,B^-_q]&=&-\omega_qB^-_q S^z_0+ S^-_0 X_q,~q=Q+1,\cdots,M\nonumber\\\\
\label{Xq2}
X_q&\equiv&\sum^N_{j=1}A_{qj} g^*_jS^z_j-A_{q0}h+A_{q0}\omega_q S^z_0.
\end{eqnarray}
The two operators $\tilde{X}_q$ and $X_q$ do not induce spin flipping and satisfy
\begin{eqnarray}
\tilde{X}_q|F\rangle&=&\tilde{x}_q|F\rangle,\\
X_q|F\rangle&=&x_q|F\rangle,
\end{eqnarray}
with eigenvalues
\begin{eqnarray}
\label{xq1}
\tilde{x}_q&=&\sum^N_{j=1} \tilde{A}_{qj} g^*_js_j,\\
\label{xq2}
x_q&=&\sum^N_{j=1}A_{qj} g^*_js_j-A_{q0}h+A_{q0}\omega_q s_0.
\end{eqnarray}
\par By noting that $H|F\rangle=0$ and using the usual trick~\cite{wu2018}, we obtain
\begin{eqnarray}\label{LHS_wq1}
H|\Psi_M\rangle&=&-s_0\sum^M_{q=Q+1}\omega_q|\Psi_M\rangle\nonumber\\
&& +\sum^Q_{q=1}\tilde{x}_qP^{M,(q)}_1|F\rangle+\sum^M_{q=Q+1}x_qP^{M,(q)}_1|F\rangle\nonumber\\
&&-\sum^M_{q=Q+1}\omega_qP^{q}_1[S^z_0,P^M_{q+1}]|F\rangle\nonumber\\
&&+\sum^Q_{q=1}P^{q-1}_1S^-_0[\tilde{X}_q,P^M_{q+1}]|F\rangle\nonumber\\
&&+\sum^M_{q=Q+1} P^{q-1}_1S^-_0[X_q,P^M_{q+1}]|F\rangle,
\end{eqnarray}
where the commutators read
\begin{eqnarray}\label{SzPM}
[S^z_0,P^M_{q+1}]&=&-\sum^M_{p=q+1}A_{p0}P^{M,(p)}_{q+1},
\label{Cm2}
\end{eqnarray}
\begin{eqnarray}
[\tilde{X}_q,P^M_{q+1}]&=&- \sum^Q_{p=q+1} P^{p-1}_{q+1}  \sum^N_{j=1} g^*_j \tilde{A}_{qj}\tilde{A}_{pj}S^-_j  P^M_{p+1}\nonumber\\
&&- \sum^M_{p=Q+1} P^{p-1}_{q+1}  \sum^N_{j=1} g^*_j \tilde{A}_{qj}A_{pj}S^-_j  P^M_{p+1},\nonumber\\
\label{Cm2a}
\end{eqnarray}
and
\begin{eqnarray}\label{XPMq1}
[X_q,P^M_{q+1}]&=& - \sum^M_{p=q+1} A_{q0}A_{p0} \omega_q  P^{M,(p)}_{q+1}\nonumber\\
&& - \sum^M_{p=q+1} P^{p-1}_{q+1}  \sum^N_{j=1} g^*_jA_{qj}A_{pj}S^-_j  P^M_{p+1}.\nonumber\\
\label{Cm2b}
\end{eqnarray}
\par To proceed further, we demand that, for example, the product $g^*_j\tilde{A}_{qj}\tilde{A}_{pj}$ should be expressible as a linear combination of $\tilde{A}_{qj}$ and $\tilde{A}_{pj}$, and similar requirements should be imposed for $g^*_j \tilde{A}_{qj}A_{pj}$ and $g^*_j A_{qj}A_{pj}$~\cite{wu2018}.  We now look at the second term in Eq.~(\ref{XPMq1}). If $Q\leq M-2$, then we are forced to deal with the expression $\sum^N_{j=1}
g^*_jA_{qj}A_{pj}S^-_j$ with $Q+1\leq q<p\leq M$. However, from Eq.~(\ref{Ctr1b}) we have
\begin{eqnarray}
&&\sum^N_{j=1} g^*_jA_{qj}A_{pj}S^-_j =\frac{A_{q0} A_{p0}}{\omega_q\omega_p} \sum^N_{j=1} |g_j|^2g_jS^-_j,
\end{eqnarray}
which can never be made proportional to any $B^-_q$ unless $g_j$ is of the form
\begin{eqnarray}\label{gjg}
g_j=|g|e^{i\theta_j},
\end{eqnarray}
where $|g|$ is the common norm of each $g_j$ and $\theta_j$ is an arbitrary real number. This can be regarded as an extension of homogenous coupling with a local gauge transformation on the bath spins. Nevertheless, below we will assume that $\{g_j\}$ are generally inhomogeneous and discuss the case given by Eq.~(\ref{gjg}) in Sec.~\ref{SecV}.
\par The above arguments indicate that $H$ is \emph{possibly} solvable via the operator product state ansatz given by Eq.~(\ref{product}) only for $Q=M$ or $Q=M-1$. As we will see, these two situations exactly correspond to the dark states and bright states revealed in Ref.~\cite{XXCSM}. Before ending this section, let us discuss the case with $M=1$ to see how these two types of states emerge.
\subsection{Single-spin-excitation subspace with $M=1$}
\par As a warm up, let us first study the simple case of $M=1$. In this case there is a single spin-excitation upon the reference state $|F\rangle$ and $Q$ can be either $1$ or $0$. From Eq.~(\ref{LHS_wq1}) we have
\begin{eqnarray}\label{LHS_wq1M1Q1}
H|\Psi^{(1)}_1\rangle&=& \tilde{x}_1 S^-_0 |F\rangle
\end{eqnarray}
for $Q=1$ and
\begin{eqnarray}\label{LHS_wq1M1Q0}
H|\Psi^{(0)}_1\rangle&=&-s_0 \omega_1|\Psi^{(0)}_1\rangle+ x_1 S^-_0 |F\rangle
\end{eqnarray}
for $Q=0$, where the superscript in the states denotes the value of $Q$. We see that if we can suitably choose the parameters $\tilde{A}_{1j}$ for $Q=1$ ($A_{1j}$ for $Q=0$) such that $\tilde{x}_1=0$ ($x_1=0$), we obtain the eigenstate $|\Psi^{(1)}_1\rangle$ ($|\Psi^{(0)}_1\rangle$) with eigenenergy $E^{(1)}_1=0$ ($E^{(0)}_1=-s_0 \omega_1$).
\par From Eq.~(\ref{xq1}), the condition $\tilde{x}_1=0$ reads
\begin{eqnarray}\label{M1eq1}
\sum^N_{j=1}\tilde{A}_{1j}g^*_js_j=0.
\end{eqnarray}
The rank-nullity theorem tells us that there are $N-1$ linearly independent solutions, $\{\tilde{A}^{(\alpha)}_{1j}\}$ ($\alpha=1,2,\cdots,N-1$), to Eq.~(\ref{M1eq1}), resulting in $N-1$ degenerate zero-energy eigenstates
\begin{eqnarray}
|\Psi^{(1)}_{1,\alpha}\rangle&=&\sum^N_{j=1}\tilde{A}^{(\alpha)}_{1j}S^-_j|F\rangle\nonumber\\
&=&|s_0\rangle\sum^N_{j=1}\tilde{A}^{(\alpha)}_{1j}\sqrt{2s_j}|s_1,\cdots,s_j-1,\cdots,s_N\rangle.\nonumber\\
\end{eqnarray}
Similarly, using Eqs.~(\ref{Ctr1b}) and (\ref{xq2}) the condition $x_1=0$ can be rewritten as
\begin{eqnarray}\label{w1sol}
s_0\omega^2_1-h\omega_1-\sum^N_{j=1}|g_j|^2 s_j=0,
\end{eqnarray}
which gives two other eigenenergies
\begin{eqnarray}\label{M1Epm}
E^{(0)}_{1,\pm}&=& -s_0\omega_{1,\pm},\nonumber\\
\omega_{1,\pm}&=&\frac{h\pm\sqrt{h^2+4s_0\sum^N_{j=1} |g_j|^2s_j}}{2s_0},
\end{eqnarray}
with the (unnormalized) eigenstates given by
\begin{eqnarray}\label{Psi01pm}
|\Psi^{(0)}_{1,\pm}\rangle&=&\sum^N_{j=0}A_{1j}S^-_j|F\rangle\nonumber\\
&=&\sqrt{2s_0}|s_0-1\rangle|s_1,\cdots,s_N\rangle\nonumber\\
&&+\frac{s_0}{E^{(0)}_{1,\pm}}|s_0\rangle\sum^N_{j=1}g_j\sqrt{2s_j}|s_1,\cdots,s_j-1,\cdots,s_N\rangle.\nonumber\\
\end{eqnarray}
It is apparent that the $N-1$ zero-energy states $|\Psi^{(1)}_{1,\alpha}\rangle$ and the two states $|\Psi^{(0)}_{1,\pm}\rangle$ are respectively the dark (separable) and bright (entangled) states revealed in Ref.~\cite{XXCSM}.
\par We can similarly take the lowest state $|F'\rangle$ as the reference state to start with. The corresponding single-spin-excitation states in the sector with $L_z=-\sum^N_{j=0}s_j+1$ can be obtained by applying the collective raising operator $B'^+_q=\sum^N_{j=1}A'_{jq}S^+_j$ to $|F'\rangle$. It is easy to show that the such obtained two entangled states possess energies
\begin{eqnarray}
E'^{(0)}_{1,\pm} =( 1-2s_0)h+E^{(0)}_{1,\pm}.
\end{eqnarray}
We thus showed that all the $N+1$ eigenstates in the $M=1$ sector are given by the operator product state ansatz. Below we concentrate on the cases with $M\geq 2$.
\section{$Q=M$: separable state}\label{SecIII}
\par In this section we discuss the case of $Q=M$ (with $M\geq 2$) for which $A_{q0}=0,~\forall q$. We will show that the operator product state
\begin{eqnarray}\label{darkPsiM}
|\Psi^{(M)}_M\rangle&=&|s_0\rangle\prod^M_{q=1}\left(\sum^N_{j=1}\tilde{A}_{qj}S^-_j\right)|s_1,\cdots,s_N\rangle,
\end{eqnarray}
provides all the separable states for arbitrary $s_0\geq1/2$.
\par For $Q=M$, Eq.~(\ref{LHS_wq1}) is reduced to
\begin{eqnarray}\label{LHS_wqa}
H|\Psi^{(M)}_M\rangle&=&\sum^M_{q=1}P^{q-1}_1S^-_0[\tilde{X}_q,P^M_{q+1}]|F\rangle +\sum^M_{q=1}\tilde{x}_qP^{M,(q)}_1|F\rangle,\nonumber\\
\end{eqnarray}
with
\begin{eqnarray}\label{XPQM}
[\tilde{X}_q,P^M_{q+1}]&=&- \sum^M_{p=q+1} P^{p-1}_{q+1}  \sum^N_{j=1} g^*_j \tilde{A}_{qj}\tilde{A}_{pj}S^-_j  P^M_{p+1}. \nonumber\\
\end{eqnarray}
Remembering that the $(\tilde{A}_{qj})$'s are still arbitrary, we have a chance to appropriately choose them such that the product $g^*_j\tilde{A}_{qj}\tilde{A}_{pj}$ can be written as a linear combination of $\tilde{A}_{qj}$ and $\tilde{A}_{pj}$. We thus impose the following constraint
 \begin{eqnarray} \label{betabeta}
g^*_j\tilde{A}_{qj}\tilde{A}_{pj}= \beta_{q,p} \tilde{A}_{qj}+\beta_{p,q} \tilde{A}_{pj},~j=1,2,\cdots,N
\end{eqnarray}
by considering that the left-hand side of the above equation is symmetric with respect to the interchange of $q$ and $p$.
\par Before discussing possible explicit forms of $\beta_{p,q}$, let us assume Eq.~(\ref{betabeta}) is already satisfied. We insert Eq.~(\ref{betabeta}) into Eq.~(\ref{XPQM}) to get
\begin{eqnarray}
[\tilde{X}_q,P^M_{q+1}]&=& - \sum^M_{p=q+1} \beta_{q,p}P^{p-1}_{q}  P^M_{p+1}  - \sum^M_{p=q+1}  \beta_{p,q}  P^M_{q+1},\nonumber\\
\end{eqnarray}
which in combination with Eq.~(\ref{LHS_wqa}) results in
\begin{eqnarray}\label{Hbetabeta}
H|\Psi^{(M)}_M\rangle&=&-\sum^M_{p>q}(\beta_{q,p}P_1^{M,(p)}+\beta_{p,q}P_1^{M,(q)})|F\rangle\nonumber\\
&& +\sum^M_{p=1}\tilde{x}_pP_1^{M,(p)}|F\rangle.
\end{eqnarray}
The first term in Eq.~(\ref{Hbetabeta}) can be rearranged as
\begin{eqnarray}
&&-\sum^M_{p>q}(\beta_{q,p}P_1^{M,(p)}+\beta_{p,q}P_1^{M,(q)})|F\rangle\nonumber\\
&=&-\sum^M_{p>q} \beta_{q,p}P_1^{M,(p)}|F\rangle-\sum^M_{q>p}\beta_{q,p}P_1^{M,(p)}|F\rangle\nonumber\\
&=&  -\sum^M_{p=1}\sum_{q(\neq p)}\beta_{q,p}P_1^{M,(p)}|F\rangle,\nonumber
\end{eqnarray}
giving
\begin{eqnarray}
H|\Psi^{(M)}_M\rangle&=& \sum^M_{p=1}\left(\tilde{x}_p-\sum_{q(\neq p)}\beta_{q,p}\right)P_1^{M,(p)}|F\rangle.
\end{eqnarray}
We see that if we set
\begin{eqnarray}\label{BAEa}
\tilde{x}_p-\sum_{q(\neq p)}\beta_{q,p}=0,
\end{eqnarray}
then $|\Psi^{(M)}_M\rangle$ is an eigenstate of $H$ with zero eigenenergy.
\par We now turn to discuss the solutions of Eq.~(\ref{betabeta}). Following Ref.~\cite{wu2018}, we seek solutions with
antisymmetric $\beta_{p,q}$, i.e.
\begin{eqnarray}
\beta_{p,q}=-\beta_{q,p},
\end{eqnarray}
for which Eq.~(\ref{betabeta}) becomes
\begin{eqnarray} \label{betabeta2}
g^*_j = \beta_{q,p} \left(\frac{1}{\tilde{A}_{pj}}- \frac{1}{\tilde{A}_{qj}}\right).
\end{eqnarray}
It is easy to see that
 \begin{eqnarray}\label{Apj}
\tilde{A}_{pj}=\frac{1}{a_j-g^*_j\nu_p},~p=1,\cdots,M
\end{eqnarray}
and
\begin{eqnarray} \label{betabeta3}
\beta_{q,p}=\frac{1}{\nu_q-\nu_p},~~q,p=1,\cdots,M
\end{eqnarray}
satisfy Eq.~(\ref{betabeta2}), where $a_j$ with $j=1,\cdots,N$ are dimensionless constants depending only on $j$. They correspond to the anisotropic parameters associated with the energies $\{\epsilon_j\}$ of the bath spins in the Gaudin-type central spin problem~\cite{Stolze2007}. The $M$ parameters $\nu_q$ ($q=1,\cdots,M$) correspond to the rapidities in the Bethe ansatz language and have the dimension of inverse energy.
\par By combining Eqs.~(\ref{xq1}), (\ref{BAEa}), (\ref{Apj}), and (\ref{betabeta3}), we finally obtain the following $M$ coupled equations
\begin{eqnarray}\label{BAEafinal}
\sum^N_{j=1}\frac{g^*_js_j}{a_j-g^*_j\nu_p}-\sum^M_{q(\neq p)}\frac{1}{\nu_q-\nu_p}=0,~p=1,\cdots,M.
\end{eqnarray}
Note that the above equations are independent of $s_0$ due to the separable nature of the state (\ref{darkPsiM}). If we choose $a_j=1/g_j$ in the above equations and reinterpret $\{\nu_q\}$ as rapidities with a different dimension, we then recover the Bethe ansatz equations for the dark states presented in Ref.~\cite{XXCSM}. However, the constants $\{a_j\}$ in Eq.~(\ref{BAEafinal}) can in principle be arbitrarily chosen [except for those rendering Eq.~(\ref{BAEafinal}) unsolvable]. This freedom of choice for $\{a_j\}$ is consistent with the fact that the separable states generally form a degenerate manifold in the $M$-sector. Different choices of $\{a_j\}$ account for different linear combinations of a fixed set of separable states. In spite of the appearance of the free parameters $\{a_j\}$, we will still refer to the $M$ coupled equations given by (\ref{BAEafinal}) as the Bethe ansatz equations, a particular form of the Bethe ansatz equations for general central spin problems~\cite{Stolze2007}. Obviously, the state $|\Psi^{(M)}_M\rangle$ should be independent of $a_j$ if it is nondegenerate in the $M$-sector.
\par As an example, let us consider the case with $M=2$ and $N=2$, for which the two Bethe ansatz equations read (we assume both $g_1$ and $g_2$  are real)
\begin{eqnarray}
\label{BAM2N2a}
 \frac{g_1s_1 }{a_1-g_1\nu_1}+ \frac{g_2s_2 }{a_2-g_2\nu_1}- \frac{1}{\nu_2-\nu_1}&=&0,\\
 \label{BAM2N2b}
 \frac{g_1s_1 }{a_1-g_1\nu_2}+ \frac{g_2s_2 }{a_2-g_2\nu_2}- \frac{1}{\nu_1-\nu_2}&=&0.
\end{eqnarray}
If we further choose $s_1=s_2=1$, it is easy to solve the above two equations to obtain a unique solution (regardless of the order of $\nu_1$ and $\nu_2$)
\begin{eqnarray}\label{v1v2}
\nu_1&=&\frac{3 (g_1a_2+g_2a_1)+i\sqrt{3} (g_1a_2-g_2a_1)}{6 g_1g_2 },\nonumber\\
\nu_2&=&\frac{3 (g_1a_2+g_2a_1)-i\sqrt{3} (g_1a_2-g_2a_1)}{6 g_1g_2 },
\end{eqnarray}
indicating that the separable state in the $M=2$ sector is actually nondegenerate. In turn, the four coefficients $\tilde{A}_{qj}$ are given by
\begin{eqnarray}\label{AAAA}
\tilde{A}_{11}&=&\frac{ \sqrt{3}i(1+\sqrt{3}i)g_2}{2(a_2g_1-a_1g_2)},~~\tilde{A}_{12}=\frac{\sqrt{3}i(1-\sqrt{3}i)g_1}{2(a_2g_1-a_1g_2)},\nonumber\\
\tilde{A}_{21}&=&-\frac{\sqrt{3}i(1-\sqrt{3}i)g_2}{2(a_2g_1-a_1g_2)},~~\tilde{A}_{22}=-\frac{\sqrt{3}i(1+\sqrt{3}i)g_1}{2(a_2g_1-a_1g_2)}.\nonumber\\
\end{eqnarray}
Thus, we indeed obtain a unique separable state
\begin{eqnarray}
|\Psi^{(2)}_{2}\rangle&=&|s_0\rangle[g^2_2(S^-_1)^2+g^2_1(S^-_2)^2-g_1g_2S^-_1S^-_2]|1,1\rangle,\nonumber
\end{eqnarray}
in the $M=2$ sector, which is independent of $a_1$ and $a_2$. The form of Eq.~(\ref{AAAA}) also suggests that the Bethe ansatz equations given by Eqs.~(\ref{BAM2N2a}) and (\ref{BAM2N2b}) are actually unsolvable for $a_j=g_j/c$, where $c$ is a constant having the dimension of energy. Unlike the case of the XXZ-type coupling for which the Bethe ansatz equations exclude the case of homogeneous coupling~\cite{Guan2019} (usually the Bethe ansatz equations for inhomogeneous XXZ-type coupling provide $C_{N+1}^{M}$ sets of solutions corresponding to all the states in the sector of $M$ down spins), i.e., they do not yield a complete set of solutions~\cite{BAbreak}, here Eqs.~(\ref{BAM2N2a}) and (\ref{BAM2N2b}) still admit solutions for a homogeneous coupling with $g_1=g_2$, provided we choose $a_1\neq a_2$. In fact, as we will show in Sec.\ref{SecV}, the homogeneous XX central spin model is indeed solvable for $s_0=1/2$.
\par For a given $M\geq 2$, we can in principle obtain all the separable states in the form of Eq.~(\ref{darkPsiM}) by solving the Bethe ansatz equations given by Eq.~(\ref{BAEafinal}). However, as mentioned in Ref.~\cite{XXCSM}, the Bethe ansatz equations (\ref{BAEafinal}) do not always admit solutions. In addition, the eigenstates given by Eq.~(\ref{darkPsiM}) are all constructed based on the highest state $|F\rangle$, and hence cannot cover those separable states with the central spin in its lowest state $|-s_0\rangle$. From symmetry considerations, the latter type of separable states also exist and can actually be constructed by choosing the lowest state $|F'\rangle$ as the reference state and $B'^+_q=\sum^N_{j=1}\tilde{A}'_{qj}S^+_j$ as the collective raising operator. The resultant separable states will be in the form of $|\Psi'^{(M')}_{M'}\rangle=|-s_0\rangle\prod^{M'}_{q=1}\left(\sum^N_{j=1}\tilde{A}'_{qj}S^+_j\right)|-s_1,\cdots,-s_N\rangle$ and possess eigenenergy $E'^{(M')}_{M'}=-2s_0h$.
\par The analysis in this section shows that the operator product state ansatz (\ref{product}) can give all the separable states for arbitrary $s_0\geq1/2$. Furthermore, the total number of separable states generated from $|F\rangle$ and $|F'\rangle$ are the same. 
\section{$Q=M-1$: Entangled states}\label{SecIV}
We now study the second possibility with $Q=M-1$, for which Eq.~(\ref{LHS_wq1}) is reduced to
\begin{eqnarray}\label{HpsiQM1}
H|\Psi^{(M-1)}_M\rangle&=&-s_0 \omega_M|\Psi^{(M-1)}_M\rangle\nonumber\\
&& +\sum^{M-1}_{q=1}\tilde{x}_qP^{M,(q)}_1|F\rangle+x_M P^{M-1}_1S^-_0|F\rangle\nonumber\\
&&+\sum^{M-1}_{q=1}P^{q-1}_1S^-_0[\tilde{X}_q,P^M_{q+1}]|F\rangle,
\end{eqnarray}
with
\begin{eqnarray}\label{XqPb}
[\tilde{X}_q,P^M_{q+1}]&=&- \sum^{M-1}_{p=q+1} P^{p-1}_{q+1}  \sum^N_{j=1} g^*_j \tilde{A}_{qj}\tilde{A}_{pj}S^-_j  P^M_{p+1}\nonumber\\
&&-  P^{M-1}_{q+1}  \sum^N_{j=1} g^*_j \tilde{A}_{qj}A_{Mj}S^-_j.
\end{eqnarray}
A general eigenstate in the $M$ sector (with $M\geq 2$) reads
\begin{eqnarray}\label{brightPsiM}
&&|\Psi^{(M-1)}_M\rangle\nonumber\\
&=&\sqrt{2s_0}|s_0-1\rangle \prod^{M-1}_{q=1}\left(\sum^N_{j=1}\tilde{A}_{qj}S^-_j\right)|s_1,\cdots,s_N\rangle\nonumber\\
&&-|s_0\rangle\left(\sum^N_{j=1}\frac{g_j}{\omega_M}S^-_j\right)\prod^{M-1}_{q=1}\left(\sum^N_{j=1}\tilde{A}_{qj}S^-_j\right)|s_1,\cdots,s_N\rangle,\nonumber\\
\end{eqnarray}
which is obviously an entangled state between the central spin and the spin bath. However, due to the restriction of the value of $Q$, we are unable to construct entangled states involving lower states $|s_0-m\rangle$ ($m\geq 2$) of the central spin, which leads to fact that the states given by Eq.~(\ref{brightPsiM}) cannot provide all the entangled states for $s_0>1/2$.
\subsection{$M=2$}
\par  Let us first study the case of $M=2$, which is actually nontrivial, as we will see. In this case the first term on the right-hand side of Eq.~(\ref{XqPb}) vanishes. From Eqs.~(\ref{HpsiQM1}) and (\ref{XqPb}) we have
\begin{eqnarray}\label{Psi12}
H|\Psi^{(1)}_2\rangle&=&S^-_0\sum^N_{j=1}(\tilde{x}_1A_{2j}+x_2\tilde{A}_{1j}-g^*_j \tilde{A}_{1j}A_{2j})S^-_j|F\rangle\nonumber\\
&&+\tilde{x}_1A_{20}(S^-_0)^2|F\rangle -s_0 \omega_2|\Psi^{(1)}_2\rangle.
\end{eqnarray}
Due to the presence of the term $\tilde{x}_1A_{20}(S^-_0)^2|F\rangle$, we have to distinguish two situations.
\subsubsection{$s_0>1/2$}
\par In this case, to achieve an eigenstate $|\Psi^{(1)}_2\rangle$ with eigenenergy $E^{(1)}_2=-s_0\omega_2$, we must set
 \begin{eqnarray}
\tilde{x}_1=0
\end{eqnarray}
since $(S^-_0)^2\neq 0$. To eliminate the first term in Eq.~(\ref{Psi12}), we further let
 \begin{eqnarray}
x_2 -g^*_j  A_{2j}=0,~\forall j
\end{eqnarray}
which can be recast as
\begin{eqnarray}\label{w2gj}
\omega^2_2s_0- h\omega_2 +|g_j|^2-\sum^N_{l=1}|g_l|^2s_l=0,~\forall j.
\end{eqnarray}
by using Eqs.~(\ref{Ctr1b}) and (\ref{xq2}). The above equations imply that the state $|\Psi^{(1)}_2\rangle$ is generally not an eigenstate of $H$ unless condition (\ref{gjg}) is satisfied.
\begin{figure}
\includegraphics[width=.48\textwidth]{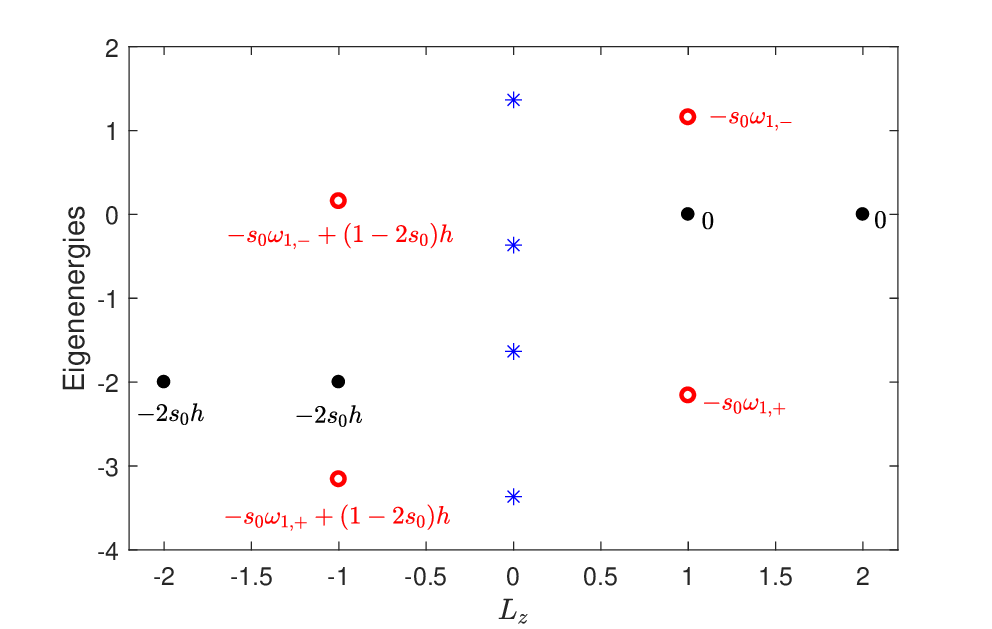}
(a)
\includegraphics[width=.48\textwidth]{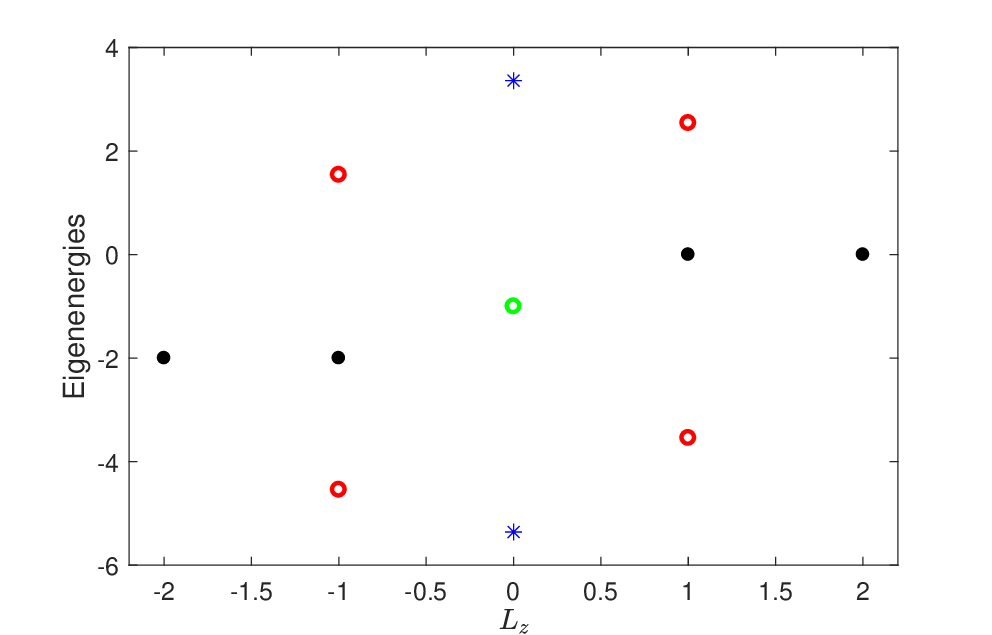}
(b)
\caption{(a) The 12 eigenenergies and the corresponding total magnetization $L_z$ for a spin-1 XX central spin model with $N=2$, $s_0=1$, $s_1=s_2=1/2$, $g_1=1$, $g_2=2$, and $h=1$. The two separable states $|\Psi^{(1)}_1\rangle$  and $|F\rangle$ (entangled states $|\Psi^{(0)}_{1,\pm}\rangle$ [see Eq.~(\ref{Psi01pm})]) are denoted by the two solid black circles (two red circles) on the right. The four blue stars in the $L_z=0$ (or $M=2$) sector correspond to the eigenstates that cannot be covered by the operator product state ansatz. The left half of the spectrum corresponds to the eigenstates constructed from $|F'\rangle$. (b) Same as Fig.~\ref{Fig1}(a), but for $g_1=g_2=3$. The green circle corresponds to the operator product state $|\Psi^{(1)}_2\rangle$ whose eigenenergy is given by Eq.~(\ref{uniformgw2}).}
\label{Fig1}
\end{figure}
\par Figure \ref{Fig1}(a) shows the energy spectrum of $H$ for $N=2$, $s_0=1$, $s_1=s_2=1/2$, $g_1=1$, $g_2=2$, and $h=1$. Since the Bethe ansatz equations given by Eqs.~(\ref{BAM2N2a}) and (\ref{BAM2N2b}) do not have any solution for $s_1=s_2=1/2$, there is only one separable state in the $M=1$ sector (solid black circle with $L_z=1$). There are also two entangled states in this sector (two red circles with $L_z=1$) whose energies are given by Eq.~(\ref{M1Epm}). The left half of the spectrum corresponds to the counterparts of the above states constructed from $|F'\rangle$. However, the four states in the $M=2$ sector (blue stars) are not included in the operator product state ansatz since $|\Psi^{(1)}_2\rangle$ is not an eigenstate of $H$ for $s_0=1$ and inhomogeneous $\{g_j\}$, as shown above.
\par For couplings having the form of $g_j=|g|e^{i\theta_j}$, we get two additional operator product states possessing energies given by solutions of Eq.~(\ref{w2gj})
\begin{eqnarray}\label{uniformgw2}
E^{(1),\mathrm{(hom)}}_{2,\pm}=-\frac{1}{2}\left[h\pm\sqrt{h^2+4s_0|g|^2\left(\sum^N_{l=1}s_l-1\right)}\right],\nonumber\\
\end{eqnarray}
which are real for $N\geq 2$. Figure \ref{Fig1}(b) shows the energy spectrum of $H$ for $N=2$, $s_0=1$, $s_1=s_2=1/2$, $g_1=g_2=3$, and $h=1$. Two degenerate eigenstates in the $L_z=0$ sector, $|\Psi^{(1)}_2\rangle$ and $|\Psi'^{(1)}_2\rangle$, appear and possess energy $E^{(1),\mathrm{(hom)}}_{2,+}=-h$ given by Eq.~(\ref{uniformgw2}) (note that $E^{(1),\mathrm{(hom)}}_{2,-}=0$ is not a physical solution since it gives $A_{20}=0$). However, there are still two states that cannot be expressed in the form of the operator product state ansatz (the two blue stars).
\par The observations in the above two examples indicate that the XX central spin model is only partially solvable for a central spin with $s_0>1/2$, i.e., the operator product state ansatz can only give the separable states and those entangled states in the single-spin-excitation sector. As we will see below, this is actually the case for all $M\geq 2$.
\subsubsection{$s_0=1/2$}
\par For $s_0=1/2$ we have $(S^-_0)^2=0$ and the solvability condition becomes $\tilde{x}_1A_{2j}+x_2\tilde{A}_{1j}-g^*_j \tilde{A}_{1j}A_{2j}=0$. Using the relation $A_{2j}=-g_jA_{20}/\omega_2$, we can recast it as
 \begin{eqnarray}\label{gjA1j}
\tilde{A}_{1j}=\frac{\tilde{x}_1g_j}{|g_j|^2+\frac{x_2\omega_2}{A_{20}}},~j=1,2,\cdots,N
\end{eqnarray}
We can use Eq.~(\ref{xq1}) to eliminate $\tilde{x}_1$ and get
\begin{eqnarray}
\sum^N_{j=1}\frac{|g_j|^2s_j}{|g_j|^2+\frac{x_2\omega_2}{A_{20}}}=1.
\end{eqnarray}
Using Eq.~(\ref{xq2}), we can rewrite the above equation as
\begin{eqnarray}\label{BAM2bb}
\sum^N_{j=1}\frac{|g_j|^2s_j}{|g_j|^2-\sum^N_{l=1}|g_l|^2s_l- h\omega_2+ \omega^2_2/2}=1.
\end{eqnarray}
Solving Eq.~(\ref{BAM2bb}) gives $K$ ($K\leq 2N$) real solutions $\omega_{2,\alpha}$ ($\alpha=1,2,\cdots,K$), and hence the eigenenergies $E_{2,\alpha}=-s_0\omega_{2,\alpha}$. The obtained $\omega_{2,\alpha}$ can then be used in the coupled linear equations given by (\ref{gjA1j}) to get the corresponding coefficients $\{\tilde{A}^{(\alpha)}_{1j}\}$. For example, for $N=2$ and $s_1=s_2=1$ the four solutions of Eq.~(\ref{BAM2bb}) read
\begin{eqnarray}\label{w1234}
\omega_{2,1}&=&1-\sqrt{1+2(|g_1|^2+|g_1g_2|+|g_2|^2)},\nonumber\\
\omega_{2,2}&=&1-\sqrt{1+2(|g_1|^2-|g_1g_2|+|g_2|^2)},\nonumber\\
\omega_{2,3}&=&1+\sqrt{1+2(|g_1|^2-|g_1g_2|+|g_2|^2)},\nonumber\\
\omega_{2,4}&=&1+\sqrt{1+2(|g_1|^2+|g_1g_2|+|g_2|^2)}.
\end{eqnarray}
\begin{figure}
\includegraphics[width=.48\textwidth]{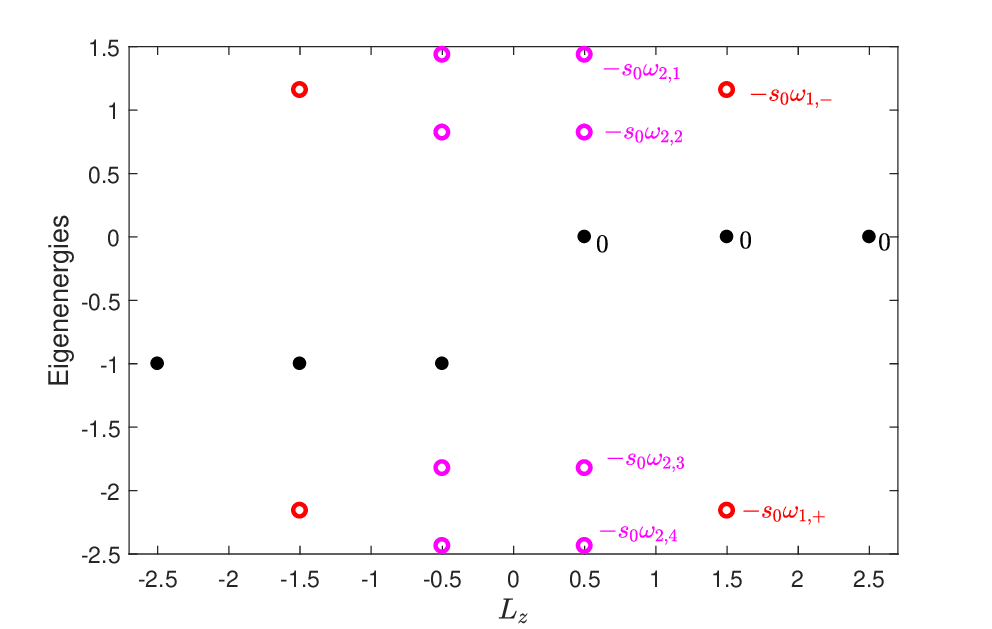}
\caption{The 18 eigenenergies and the corresponding total magnetization $L_z$ for a spin-1/2 XX central spin model with $N=2$, $s_0=1/2$, $s_1=s_2=1$, $g_1=1$, $g_2=2$, and $h=1$. The eigenenergies of the four entangled states in the $M=2$ sector (pink circles on the right) are given by Eq.~(\ref{w1234}).}
\label{Fig2}
\end{figure}
\par Figure ~\ref{Fig2} shows the spectrum of $H$ for $N=2$, $s_0=1/2$, $s_1=s_2=1$, $g_1=1$, $g_2=2$, and $h=1$. It can be seen that all the eigenstates are given by the operator product state (\ref{product}), meaning that the XX central spin model with $s_0=1/2$ is indeed solvable in the $M=2$ sector~\cite{XXCSM}.
\par It is interesting to note that Eq.~(\ref{BAM2bb}) reduces to a quadratic equation for couplings of the form $g_j=|g|e^{i\theta_j}$:
\begin{eqnarray}\label{BAM2bbhom}
|g|^2\left(1-2\sum^N_{l=1}s_l\right)- h\omega_2+ \frac{1}{2}\omega^2_2=0,
\end{eqnarray}
yielding only two solutions
\begin{eqnarray}\label{uniformgw3}
\tilde{E}^{(1),\mathrm{(hom)}}_{2,\pm}=-\frac{1}{2}\left[h\pm\sqrt{h^2+2|g|^2\left(2\sum^N_{l=1}s_l-1\right)}\right].
\end{eqnarray}
\subsection{$M\geq 3$}
\par For $M\geq 3$ we have to deal with both the two terms appearing in Eq.~(\ref{XqPb}). Following the analysis in Sec.~\ref{SecIII}, we may choose
\begin{eqnarray}\label{Aqjbright}
\tilde{A}_{qj}=\frac{1}{a_j-g^*_j\nu_q},~ q=1,\cdots,M-1
\end{eqnarray}
and
\begin{eqnarray}
\beta_{q,p}=\frac{1}{\nu_q-\nu_p},~~q,p=1,\cdots,M-1
\end{eqnarray}
such that
\begin{eqnarray}\label{AAvv}
g^*_j\tilde{A}_{qj}\tilde{A}_{pj}=\frac{\tilde{A}_{qj}-\tilde{A}_{pj}}{\nu_q-\nu_p}.
\end{eqnarray}
We note that Eq.~(\ref{gjA1j}) (in the case of $M=2$) also has the form of Eq.~(\ref{Aqjbright}) with
\begin{eqnarray}\label{ajnu1M2}
a_j&=&\frac{x_2\omega_2}{g_jA_{20}\tilde{x}_1},~~\nu_1=-\frac{1}{\tilde{x}_1}.
\end{eqnarray}
As to the second term in Eq.~(\ref{XqPb}), we wish to find two coefficients $\gamma_{M,q}$ and $\theta_{M,q}$ such that
\begin{eqnarray}\label{AAqM}
 g^*_j \tilde{A}_{qj}A_{Mj}&=& \gamma_{M,q}\tilde{A}_{qj}+\theta_{M,q}A_{Mj}.
\end{eqnarray}
Using the explicit form of $\tilde{A}_{qj}$ and the relation $A_{Mj}=-g_jA_{M0}/\omega_M$, the above condition can be reexpressed as
\begin{eqnarray}\label{abovea}
A_{M,0}g_j[\theta_{M,q}(a_j-g^*_j\nu_q)-g^*_j]&=&\gamma_{M,q}\omega_M.
\end{eqnarray}
Note that the right-hand side of Eq.~(\ref{abovea}) is independent of $j$, the parameters $\{a_j\}$ must be chosen so as to make the left-hand side independent of $j$ as well. Depending on whether $\gamma_{M,q}$ vanishes, there exist two possible choices:
\par a) If $\gamma_{M,q}=0$, we have $a_j=g^*_j(1/\theta_{M,q}+\nu_q)$ from $A_{M0}g_j\neq 0$. However, such a choice is actually unphysical due to the absence of $\gamma_{M,q}$. More importantly, it can be checked that the such obtained Bethe ansatz equations do not admit any solution.
\par b) If $\gamma_{M,q}\neq 0$, we then have
\begin{eqnarray}
\theta_{M,q}a_jg_j-|g_j|^2(\theta_{M,q}\nu_q+1)=\gamma_{M,q}\omega_M/A_{M0}.
\end{eqnarray}
By noting that the above equation must hold for any $j$, we must have
\begin{eqnarray}\label{thetaMq}
 \theta_{M,q}&=&-\frac{1}{\nu_q},
\end{eqnarray}
and hence
\begin{eqnarray}\label{ajcgj1}
a_j=-(\nu_q\gamma_{M,q})\frac{\omega_M}{A_{M0}g_j}.
\end{eqnarray}
The above equation indicates that $a_j$ should be proportional to $1/g_j$, which is in consistent with Eq.~(\ref{ajnu1M2}) for $M=2$. However, equation (\ref{ajcgj1}) must hold for any $q$, indicating that $\gamma_{M,q}$ should be proportional to $1/\nu_q$ with a $q$-independent coefficient. We thus let
\begin{eqnarray} \label{rMq}
\gamma_{M,q}=-\frac{A_{M0}}{\nu_q\omega_M}c,
\end{eqnarray}
where $c$ is a constant having the dimension of energy. Applying Eq.~(\ref{rMq}) in Eq.~(\ref{ajcgj1}) gives
\begin{eqnarray}\label{ajcgj}
a_j=\frac{c}{g_j}.
\end{eqnarray}
\par Combining Eqs.~(\ref{HpsiQM1}), (\ref{XqPb}), (\ref{AAvv}), (\ref{AAqM}), (\ref{thetaMq}) and (\ref{rMq}) and after a straightforward calculation, we obtain
\begin{eqnarray}\label{HpsiQM3}
&&H|\Psi^{(M-1)}_M\rangle=- \sum^{M-1}_{q=1}\frac{A_{M0}}{ \nu_q} P^{q-1}_1S^-_0 P^{M-1}_{q+1}S^-_0|F\rangle\nonumber\\
&&~~ +\sum^{M-1}_{p=1}\left(\tilde{x}_p+\frac{1}{\nu_p}-\sum_{q(\neq p)}\frac{1}{\nu_q-\nu_p}\right)P_1^{M,(p)} |F\rangle\nonumber\\
&&~~ +\left(x_M +\frac{A_{M,0}}{\omega_M}\sum^{M-1}_{q=1}   \frac{c}{\nu_q}\right)P^{M-1}_1S^-_0|F\rangle\nonumber\\
&&~~-s_0 \omega_M|\Psi^{(M-1)}_M\rangle.
\end{eqnarray}
Similar to the case of $M=2$, the first term on the right-hand side of the above equation can only be eliminated for $s_0=1/2$ since $A_{M0}\neq 0$ by assumption. In other words, the eigenstates of $H$ for $s_0>1/2$ and $M\geq 2$ cannot be covered by the operator product state ansatz. This means that $H$ is only \emph{partially solvable} for $s_0>1/2$, in the sense that only those entangled states in the $M=1$ sector are given by the operator product state ansatz. For $s_0=1/2$, it is shown in Ref.~\cite{XXCSM} that the operator product states indeed provide a complete set.
\par Below we will set $s_0=1/2$. The elimination of the second term in Eq.~(\ref{HpsiQM3}) yields the Bethe ansatz equations
\begin{eqnarray}\label{BAEbright}
\sum^N_{j=1}\frac{|g_j|^2s_j}{c-|g_j|^2\nu_p}+\frac{1}{\nu_p}-\sum^{M-1}_{q(\neq p)}\frac{1}{\nu_q-\nu_p}=0
\end{eqnarray}
by using
\begin{eqnarray}
\tilde{A}_{pj}=\frac{g_j}{c-|g_j|^2\nu_p}.
\end{eqnarray}
By further eliminating the third term in Eq.~(\ref{HpsiQM3}), we arrive at
 \begin{eqnarray}
&&-\sum^N_{j=1} |g_j|^2s_j- \omega_Mh+ \frac{1}{2}\omega_M^2 +\sum^{M-1}_{q=1}   \frac{ c}{\nu_q}=0,
\end{eqnarray}
which gives two branches of eigenenergies
 \begin{eqnarray}\label{EM}
E^{(M-1)}_{M,\pm}=-\frac{1}{2}\left[ h\pm\sqrt{ h^2+2\left(\sum^N_{j=1} |g_j|^2s_j -\sum^{M-1}_{q=1}   \frac{c}{\nu_q}\right )}\right].\nonumber\\
\end{eqnarray}
Note that Eqs.~(\ref{BAEbright}) and (\ref{EM}) recover the results in Ref.~\cite{XXCSM} if we set $c=1$ and reinterpret $\{\nu_q\}$ as rapidities with a different dimension. Actually, the results (the eigenstatates and eigenenergies) are independent of the constant $c$ since we can perform the rescaling $\nu_q\to c\nu_p$ to obtain $c$-independent expressions.
\section{The case of $g_j=|g|e^{i\theta_j}$}\label{SecV}
\par In this section we will study the special case given by Eq.~(\ref{gjg}), so that the Hamiltonian becomes
\begin{eqnarray}\label{Hxxcom}
H_{\mathrm{hom}}&=&h(S^z_0-s_0)+\frac{|g|}{2}\sum^N_{j=1}(e^{i\theta_j} S^+_0S^-_j+e^{-i\theta_j}S^-_0S^+_j).\nonumber\\
\end{eqnarray}
By defining $\tilde{S}^-_j=e^{i\theta_j}S^-_j,~\tilde{S}^+_j=e^{-i\theta_j}S^+_j$, and $\tilde{S}^z_j=S^z_j$ that preserve the canonical commutation relations of spins, we see that $H_{\mathrm{hom}}$ is actually equivalent to an XX central spin model with homogeneous coupling $|g|$ by a local gauge transformation on the bath spins. Such a correspondence indicates that the spectrum of $H_{\mathrm{hom}}$ is independent of the phase factors $\{e^{i\theta_j}\}$, which nevertheless enter the explicit expressions of the eigenstates.
\par The homogeneous XX central spin model has been widely studied for $s_1=\cdots=s_N=1/2$, where the collective bath lowering operator can be treated as a large spin $\tilde{S}^-=\sum^N_{j=1}\tilde{S}^-_j$. The XXZ central spin model with homogeneous coupling and $s_1=\cdots=s_N=1/2$ and arbitrary $s_0$ has been solved in Ref.~\cite{Guan2018}. The aim of this section is to show that $H_{\mathrm{hom}}$ with $s_0=1/2$ and arbitrary $\{s_j\}$ is solvable and admits operator product state ansatz solutions.
\par Our starting point is Eq.~(\ref{LHS_wq1}), in which the commutators can be calculated under condition (\ref{gjg}) as
\begin{eqnarray}
[\tilde{X}_q,P^M_{q+1}]&=&- \sum^Q_{p=q+1} P^{p-1}_{q+1}  \sum^N_{j=1} g^*_j \tilde{A}_{qj}\tilde{A}_{pj}S^-_j  P^M_{p+1}\nonumber\\
&&+ |g|^2\sum^M_{p=Q+1}\frac{A_{p0} }{\omega_p}P^{p-1}_{q}  P^M_{p+1},\\
~[X_q,P^M_{q+1}]&=& -A_{q0}\left(\frac{|g|^2 }{\omega_q}+\omega_q\right)\sum^M_{p=q+1} A_{p0}P^{M,(p)}_{q+1}\nonumber\\
&& +\frac{A_{q0} }{\omega_q}|g|^2(M-q) P^M_{q+1},
\end{eqnarray}
where $\tilde{A}_{qj}$ is given by Eq.~(\ref{Apj}). By inserting the above two equations and Eq.~(\ref{SzPM}) into Eq.~(\ref{LHS_wq1}) and after some manipulation, we get
\begin{eqnarray}\label{HpsiMhom}
&&H|\Psi^{(Q)}_M\rangle=-s_0\sum^M_{q=Q+1}\omega_q|\Psi^{(Q)}_M\rangle\nonumber\\
&&~+\sum^Q_{p=1}\left(\tilde{x}_p-\sum^Q_{q(\neq p)}\beta_{q,p}\right)P^{M,(p)}_1|F\rangle\nonumber\\
&&~+\sum^M_{p=Q+1}\left[|g|^2\frac{A_{p0} }{\omega_p}(M+Q-p)+x_p \right]P^{M,(p)}_1|F\rangle\nonumber\\
&&~+\sum^M_{p=Q+1}\sum^M_{q=p+1}\omega_pA_{q0} P^{M,(q)}_{1}|F\rangle\nonumber\\
&&~-\sum^M_{q=Q+1} \sum^M_{p=q+1}A_{q0} A_{p0} \frac{|g|^2+\omega^2_q }{\omega_q} P^{q-1}_1S^-_0P^{M,(p)}_{q+1}|F\rangle,\nonumber\\
\end{eqnarray}
where $\beta_{p,q}$ is given by Eq.~(\ref{betabeta3}). We note that for the homogenous coupling \emph{all possible values} of $Q$ (with $0\leq Q\leq M$) are allowed.
However, this does not mean that all the entangled state involving $|s_0-m\rangle$ with $m\geq 2$ can be given by the operator product state ansatz. We already see this from the simple example shown in Fig.~\ref{Fig1}(b) for $s_0=1$, where only some of the states in the $M=2$ sector (the green circle) are given by the operator product state ansatz. In fact, to eliminate the last term in Eq.~(\ref{HpsiMhom}) we are still forced to set $s_0=1/2$.
\par To proceed, we assume $s_0=1/2$ and note that
\begin{eqnarray}
&&\sum^M_{p=Q+1}\sum^M_{q=p+1}\omega_pA_{q0} P^{M,(q)}_{1}|F\rangle\nonumber\\
&=&\sum^M_{p=Q+2}\sum^{p-1}_{q=Q+1}\omega_qA_{p0} P^{M,(p)}_{1}|F\rangle,
\end{eqnarray}
which converts Eq.~(\ref{HpsiMhom}) to
\begin{eqnarray}\label{HpsiMhom1}
&&H|\Psi^{(Q)}_M\rangle=-\frac{1}{2}\sum^M_{q=Q+1}\omega_q|\Psi^{(Q)}_M\rangle\nonumber\\
&&~+\sum^Q_{p=1}\left(\tilde{x}_p-\sum^Q_{q(\neq p)}\beta_{q,p}\right)P^{M,(p)}_1|F\rangle\nonumber\\
&&~+\sum^M_{p=Q+1}\left[|g|^2\frac{A_{p0} }{\omega_p}(M+Q-p)+x_p \right]P^{M,(p)}_1|F\rangle\nonumber\\
&&~+\sum^M_{p=Q+2}\sum^{p-1}_{q=Q+1}\omega_qA_{p0} P^{M,(p)}_{1}|F\rangle.
\end{eqnarray}
Letting the second line of the above equation be zero, we arrive at the $Q$ coupled Bethe ansatz equations
\begin{eqnarray}\label{BAEuniform}
\sum^N_{j=1}\frac{|g| s_j}{a_je^{i\theta_j}-|g| \nu_p}-\sum^{Q}_{q(\neq p)}\frac{1}{\nu_q-\nu_p}=0,~p=1,\cdots,Q
\end{eqnarray}
which have the same form as Eq.~(\ref{BAEafinal}) for the separable states by using $g^*_j=|g|e^{-i\theta_j}$. As mentioned above, the phase factors $\{e^{i\theta_j}\}$ do enter the wave functions through the rapidities $\{\nu_q\}$. We can always choose suitable $\{a_j\}$ to guarantee the existence of physical solutions of Eq.~(\ref{BAEuniform}). Similar to the case of the separable states for the inhomogeneous coupling, this freedom of choice of $\{a_j\}$ also indicates that some of the eigenstates of the homogeneous model could be highly degenerate for large enough $Q$. Actually, by eliminating the last two lines of Eq.~(\ref{HpsiMhom1}), we arrive at the following $M-Q$ coupled equations
\begin{eqnarray}
&&|g|^2 \left(M+Q-p-\sum^N_{j=1} s_j\right) - h\omega_p+ \frac{1}{2}\omega^2_p\nonumber\\
&&+ \omega_p\sum^{p-1}_{q=Q+1}\omega_q=0,~~p=Q+1,\cdots,M
\end{eqnarray}
which give the eigenenergy
\begin{eqnarray}
E^{(Q)}_M=-\frac{1}{2}\sum^M_{q=Q+1}\omega_q.
\end{eqnarray}
It is interesting to note that the eigenenergy $E^{(Q)}_M$ does not depend on the parameters $\{\nu_q\}$, but only on $|g|, h, \{s_j\}$, and $Q$, consistent with our arguement that some of the eigenstates of $H_{\mathrm{hom}}$ could be degenerate.
\section{Conclusions and Discussions}\label{Secfinal}
\par In this work, we have studied partial solvability of the XX central spin model with arbitrary central spin moment $s_0\geq1/2$. By employing the operator approach based on a commutator scheme that has been previously applied to various Gaudin-like models, we have obtained both the separable and entangled states of the XX central spin model with $s_0=1/2$ through an operator product state ansatz, confirming the results presented in a recent study~\cite{XXCSM}. The corresponding Bethe ansatz equations are derived. It is found that the Bethe ansatz equations associated with the separable states are actually non-unique due to the high degeneracy of these states.
\par In addition, we show that for $s_0>1/2$ only some of the eigenstates, i.e., all the separable states and those entangled states in the single-spin-excitation subspace admit the form of the operator product state ansatz. Finally, we found that our method can also be applied to the case of homogeneous coupling. We derive the Bethe ansatz equations determining the rapidities and a coupled system of nonlinear equations that give the eigenenergies, which are found independent of the rapidities.
\par Although the partial solvability displayed here is reminiscent of quasi-exactly solvable systems~\cite{Turb88,Uly92,Ushv93}, there are some fundamental differences and questions which should be explored in the future. In particular, it would be useful to determine whether the subspace of solvable states for $s_0>1/2$ can be understood in terms of an invariance under the action of some algebraic structure. Also, while the non-uniqueness of a Bethe ansatz solution has been seen in other contexts, e.g.~\cite{Dunn08}, the level of generality of Bethe ansatz equations found for separable states in this study is unexpected and deserves to be investigated at a deeper level.
\par The separable and entangled states were known to play an important role in the control of the mesoscopic spin bath by a central spin manipulation~\cite{Lukin2003,Cappel2013}. The separable states supported by the high-spin XX central spin model will be useful in the cooling or polarization of the spin bath through the manipulation of a central spin with large quantum number. Our concrete treatment of the XX central spin model using the operator approach and the obtained results pave the way toward finding simple solutions to Gaudin-like models. In particular, our method also offers a promising opportunity to study  nonequilibrium dynamics and quench dynamics, e.g., the quantification and real-time evolution of entanglement and Fisher information in related central spin systems.

\noindent{\bf Acknowledgements:}
This work was supported by the National Natural Science Foundation of China (NSFC) under Grant No. 11705007, and partially by the Beijing Institute of Technology Research Fund Program for Young Scholars. X.W.G. is supported by the key NSFC Grant No. 11534014 and NSFC Grant No. 11874393, and the National Key R\&D Program of China No. 2017YFA0304500. J.L. acknowledges support from the Australian Research Council through Discovery Project DP200101339.

\end{document}